\documentclass{aa}
\input{psfig}
\begin{document}
\thesaurus{03                     
            (11.09.1 NGC 2146     
             11.09.2              
             11.11.1              
	     11.19.2              
             13.19.1) }           
\title{H$I$ Observations of the starburst galaxy NGC 2146}
\author{A. Taramopoulos \inst{1}\fnmsep\inst{2} \and H. Payne \inst{3} 
\and F. H. Briggs \inst{4} }
\offprints{A. Taramopoulos}
\institute{Technological Institute of Thessaloniki, 
           Department of Electronics, 
           Al. Papanastasiou 13, Thessaloniki, 54639, Greece
          \and
           Library Information Services,
           University of Macedonia, 
           Egnatia 156, Thessaloniki, 54006, Greece \\
           email: ttar@uom.gr
          \and
           Space Telescope Science Institute, 
           3700 San Martin Drive, Baltimore, MD 21218, USA \\
           email: payne@stsci.edu
          \and
           Kapteyn Astronomical Institute, 
           Postbus 800, 9700 AV Groningen, The Netherlands \\
           email: fbriggs@astro.rug.nl
         }

\date{Received 18 May 2000 ; accepted 16 October 2000 }

\maketitle

\begin{abstract}

NGC~2146 is a peculiar spiral galaxy which is currently undergoing a
major burst of star formation and is immersed in a extended H\,I
structure that has morphological and kinematical resemblence to
a strong tidal interaction.  This paper reports
aperture synthesis observations carried out in the 21cm line
with the Very Large Array (VLA\footnote{The National Radio Astronomy 
Observatory (NRAO) is
operated by Associated Universities, Inc. under cooperative agreement with
the National Science Foundation.}) of two fields
positioned to optimally cover the H\,I streams to the north
and south of the galaxy, along with a 300 ft total power spectral
mapping program to recover the low surface brightness extended
emission.
The observations
reveal elongated streams of neutral hydrogen towards both
the north and the south of the optical galaxy extending out up
to 6 Holmberg radii. The streams are not in the principle plane of rotation 
of the galaxy, 
but instead are suggestive of a 
tidal interaction between NGC~2146 and a LSB companion that was
destroyed by the encounter and remains undetected at optical wavelengths.
Part of the southern stream is  
turning back to fall into the main galaxy, where it will create a
long-lived warp in the H\,I disk of NGC~2146.  Analysis of the trajectory
of the outlying gas suggests that the closest encounter took place 
about 0.8 billion years ago and that infall of debris will continue
for a similar time span.


\end{abstract}

\section{Introduction}

NGC~2146 is a peculiar spiral galaxy as seen from both 
the optical image and the H\,I line profile. 
Although it is classified as Sab by de Vaucouleurs et~al. 
(\cite{dev76}), it shows a broad range of peculiarities. Measured by its far 
infrared flux, it is one of the 12 brightest galaxies in the sky and
lies at a distance of 12.2 Mpc ($H_0$=75~km~s$^{-1}$~Mpc$^{-1}$,
and 1$^\prime$ corresponds to 3.5 kpc). In optical images there 
are two well-defined arms which mark the principal plane of rotation. 
Superimposed across part of the nucleus is an absorption band having 
``the form of a hand, with four talon-like fingers" (Pease \cite{pease}),
which de Vaucouleurs (\cite{dev50}) interpreted as being a third arm inclined to 
the plane of rotation. Further optical studies by Benvenuti et~al. (\cite{benvenuti}),
however, suggest that a simple spiral model is not adequate. 

In 1976,
Fisher \& Tully mapped the region around this galaxy
in the 21cm
using the NRAO 91-m telescope with a resolution of $11^\prime .3  \times 10^\prime .2$ 
in the N-S and E-W directions respectively and discovered an extensive 
``envelope'' of neutral hydrogen 
around it which extends up to six Holmberg radii (100 Kpc). They suggest that the 
abnormalities seen both optically and in the neutral hydrogen profile of 
the main disk are probably related to the large H\,I extensions observed, 
and believe that they might be the result of \newline
\hspace*{0.5cm} a) tidal interaction, \newline
\hspace*{0.5cm} b) explosion/ejection, or \newline
\hspace*{0.5cm} c) galaxy formation.

They rule out any form of interaction between NGC~2146 and NGC~2146A, an 
Sc typed galaxy with no evident optical abnormalities, which lies about 
20 arcseconds away and is redshifted 595~km~s$^{-1}$ with respect to 
NGC~2146.
However, their observations mainly aimed to trace the total extent
of the H\,I cloud, and their resolution was not good enough to 
allow them to draw more definite conclusions on the causes of these 
abnormalities. Also, the appearance of the H\,I envelope as a large
gaseous halo around the main galaxy left room for speculations
as to how large galactic halos really are, and what the impact of
this might be on the QSO absorption line system statistics.

NGC~2146 contains a strong radio source, $\sim$ 3 kpc in size, within 
its nuclear region, identified with 4C 78.06 (Caswell \& Wills \cite{caswell}). 
Kronberg \& Biermann (\cite{kronberg}) used the NRAO 
interferometer and the VLA to map the radio structure of the nuclear region. 
They found that the radio center lies in the optically obscured dust lane, but
it shows no evidence of a double nucleus. The radio continuum map agrees
very well with the CO intensity map, and unlike the optical image, it shows a
remarkable degree of symmetry (Kronberg \& Biermann \cite{kronberg}; Jackson \& Ho \cite{jackson}).
The velocity curves measured in various lines in the optical and infrared
are quite regular after allowance for extinction effects due to the dust lane
(Prada et~al. \cite{prada}). Kronberg \& Biermann (\cite{kronberg}) suggested that a strong
burst of star formation is responsible for the strong radio and infrared
emission. A $^{12}$CO, $^{13}$CO and CS study undertaken by 
Xie et~al. (\cite{xie}) places the average temperature in the nuclear
region about 55 K, and the average density at 2~$\times$~10$^4$~cm$^{-3}$.
Further evidence of a burst of start formation comes from X-ray observations 
carried out by Armus et~al. (\cite{armus}) and Della Ceca et~al. (\cite{dellaceca}) which 
reveal a galactic-scale outflow of gas driven by an intensive star bursting
activity, referred to as a starburst-driven superwind. Furthermore, this 
activity can produce long-lived bending instabilities as suggested by N-body
simulations carried out by Griv \& Chiueh (\cite{griv}) to explain the snake-shaped 
radio structure observed by Zhao et~al. (\cite{zhao}) at an angular resolution of 
2$^{\prime\prime}$ using the VLA. 

Observations probing the molecular content in CO and H$_2$ as 
well as the ionized gas content (H\,II regions) were made by Young et~al. (\cite{young}). 
They found a very large
concentration of gas in the nucleus, confirming Condon et~al. (\cite{condon}) 
earlier conclusion that this galaxy has a high star-formation rate and then, 
derived a mass-to-light ratio characteristic of very young stellar 
systems. All the above led these authors to suggest that NGC~2146 has 
recently undergone a collision with some other galaxy. The existance
of an extended arc of H\,II regions encircling the central bright 
region, which exhibit velocities which are 130$-$200~km~s$^{-1}$
higher than those expected if they are rotating in the plane of the galaxy
(Young et~al. \cite{young}), might also be interpreted as evidence of a collision. 

After undertaking an optical and infrared study of the galaxy, 
Hutchings et~al. (\cite{hutchings}) found no sign of an active nucleus
but did find many signs of a significant
population of hot young stars in the central regions of the system.
They concluded that NGC~2146 is a merging system, now in its final stages.
The dominant galaxy is seen close to edge on, and the small companion
has been stripped, leaving no sign of its nucleus. They also note that such
a scenario can also be supported by numerical simulations (Barnes \cite{barnes}).

In order to better understand the nature of this system and decide amongst the 
various scenaria which have been suggested, we obtained higher resolution 
21cm maps of the H\,I distribution around NGC~2146 using the
VLA, and combined these with 21cm observations of the NRAO 91-m
telescope to recover the emission on large angular scales, which the 
interferometric observations alone are incapable of sensing.The nature
of large gaseous halos is important in the interpretation of QSO
absorption line spectra (c.f Rao \& Briggs \cite{rao}), where the large
cross sections implied by the Fisher \& Tully observations (\cite{ft76}) 
would cause this kind of galaxies to intervene frequently by chance
if they are common in the galaxy population.

\section{Observations and Reduction}

The observations presented here consist of two parts. Interferometry 
observations, carried out with the Very Large Array, and single dish 
observations, carried out with the 300ft NRAO antenna before it 
collapsed. 

On 25,26 and 29 October 1984,   
the VLA was used in its most compact 
configuration to observe two fields, centered at $\alpha_{1950} = 6^h 
12^m 00^s, \delta_{1950} = +78^o 10^{\prime} 00^{\prime \prime}$ and at 
$\alpha_{1950} = 6^h 10^m 00^s, \delta_{1950} = +78^o 35^{\prime} 
00^{\prime \prime}$ respectively. The center fields were chosen in such 
a way that their combination produces a nearly uniform response over the
area between and including the two pointing coordinates; this covers
the main body of the galaxy and substantial fraction of the extended 
H\,I emission. A 
total of 5.5 hours per field with 25 antennas with baselines ranging 
from 36~m to a little more than 1~km were used. We observed only the right 
polarization in a band centered at a heliocentric velocity of 
915~km~s$^{-1}$. After on-line Hanning smoothing, we recorded 31 independent 
channels ranging from 583 to 1226~km~s$^{-1}$ with a FWHM velocity 
resolution of 20.75~km~s$^{-1}$. 
A broader bandwidth ``continuum channel 0'' 
to  accompany the line data was calculated 
on-line from the average of 48 such channels centered at 915~km~s$^{-1}$. 
The VLA observational parameters are summarized in Table~\ref{Obspar}.
For the flux calibration we used 3C286, adopting a flux density of 
14.88~Jy according to the latest VLA measurments. The phase 
calibration source was 0836+710 for which a flux density of 4.12~Jy at 
1.416~MHz was determined. Both calibrators were used in extracting 
a measure of the shape of the passband which we applied as a 
correction to the observations of NGC~2146.

\begin{table}[htbp]
\begin{center}
\caption{The VLA Observational parameters}
\label{Obspar}
\label{een}
\smallskip
\begin {tabular}{l l}
\hline
\hline
\noalign{\medskip}
Instrument&Very Large Array (VLA)  \\
Object&NGC 2146\\
Total observing time & 11 hours  \\
Pointing center &  \\
South field R.A. (1950.0) &6$^h$12$^m$00$^s$  \\
\phantom{South field} dec. (1950.0) & +78$^{\rm o}$10$^{\prime}$00$^{\prime\prime}$  \\
North field R.A. (1950.0) &6$^h$10$^m$00$^s$  \\
\phantom{North field} dec. (1950.0) &+78$^{\rm o}$35$^{\prime}$00$^{\prime\prime}$  \\
Flux density calibrator & 3C286  \\
Phase calibrator & 0836+710  \\
Central
velocity of spectral band & 915 km s$^{-1}$  \\
Configuration& D Array \\
Observed baselines (min-max) & 40 - 1030 m  \\
Number of  antennas & 25  \\
Number of  channels & 31 \\
Channel spacing & 20.75 km s$^{-1}$  \\ 
FWHM velocity resolution$^a$ &20.75 km s$^{-1}$  \\ 
Total line bandwidth & 3 MHz   \\ 
&643.5 km s$^{-1}$  \\
Total continuum bandwidth & 4.69 MHz   \\ 
Type of weighting of {\it u - v} data&natural \\ 
R.m.s. noise per channel & 1 - 1.25 mJy \\
Synthesized Beam (N-S E-W)& 49$^{\prime\prime} \times$ 77$^{\prime\prime}$ \\
\noalign{\smallskip}
\hline
\hline
\end{tabular}
\end{center}
\noindent Notes \\
$a$. On-line Hanning taper mode.
\end{table}

To optimize the sensitivity to weak extended H\,I emission, the Fourier 
transformation was carried out using a natural weighting for the data, 
i.e. each visibility cell in the $u-v$ plane has been weighted according 
to the integration time spent on it. This gives the best signal-to-noise 
ratio for detecting
weak sources. Since $u-v$ tracks tend to spend more time per unit 
area near the $u-v$ origin, natural weighting emphasizes the data from 
short spacings, and  produces a broad synthesized beam with an
extended, low-level  sideloabe. The natural-weighted maps 
provide sensitive information on the larger scale  and are   
suitable for analyzing the structure of the extended H\,I cloud 
of NGC~2146. However, the poor synthesized beam shape means that
care must be taken in the decovolution phase of the data processing.
Incorporation of the single dish spectral is required for reliable
mapping of the extended structure.

The construction of the continuum subtracted channel maps for each of the 
two fields is a crucial step in the reduction. Due to restrictions 
in the number of spectral channels that could be recorded in 1986 
by the VLA, the correlator configuration was chosen so that it had 
H\,I emission in 29 of the 31 velocity channels. 
The on-line ``channel 0'', being 
a mean of 48 channels, 29 of which are corrupted by the strong H\,I emission
does not represent a true continuum channel.
A true continuum map was obtained according to the 
following scheme, which is a weighted difference between spectral 
channel maps with gas and the ``channel 0'' map: 
$$Cont.Map = {1\over {(48-31)}} \left( 48 (Ch.0) -
\sum_{\scriptstyle i=1}^{31} Ch.i \right) $$
In this way, the continuum map was formed from the frequency band outside the 
channels containing H\,I emission. The continuum was then 
subtracted from the dirty line channel 
maps to obtain the narrow band H\,I emission. This yielded a continuum field 
of total flux 1.83~Jy with the brightest source being the galaxy NGC~2146 
(0.75~Jy).

During September 1987 the 300ft NRAO telescope was used to observe the 
area of NGC~2146. On each day a single drift scan was performed at a fixed 
declination, and spectra were recorded in short integrations. At
the end of the observing session, a critically sampled map of the region
surrounding NGC~2146 for merging with the interferometer data
was constructed by gridding the data onto
the same spatial and velocity grids as used for the VLA data cubes. 
For the joint deconvolution performed simultaneously on the VLA and
300ft data, a 
gaussian  beam of HPBW 11.3~arcmin in the direction N-S 
and 10.2~arcmin in the direction E-W was adopted for the
300ft (Fisher \& Tully \cite{ft75} ). 

The continuum-free VLA maps were then simultaneously cleaned, mosaiced
and combined together with the single dish maps with the AIPS 
maximum entropy based task UTESS to form single maps for each channel.
At this stage, tests 
were performed using other alternative approaches of deconvolution like 
CLEAN algorithms alone, 
or combinations of CLEAN and maximum entropy based algorithms (implemented
by VTESS in AIPS), 
yielding similar results in terms of map
quality, but making apparent the fact that the single dish data were 
adding to the final maps substantial flux that the interferometry maps 
alone were missing, due to the lack of short spacing $u-v$ coverage.

The final natural-weighted channel maps have an rms noise per channel of 
1.0 to 1.25~mJy/beam, which is approximately the theoretical 
value (1.0~mJy/beam). The final synthesized beam is 49$^{\prime\prime} 
\times $77$^{\prime\prime}$ in the North-South and East-West directions 
respectively.

\begin{figure}
\vspace{0cm}
\hspace{0cm}\psfig{figure=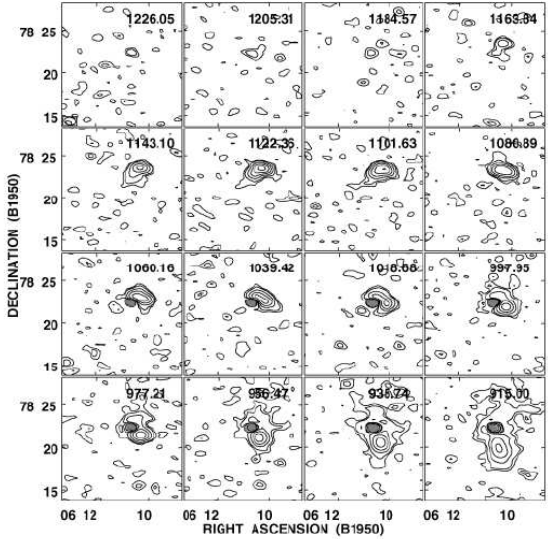,width=8.5cm}
\vspace{0cm}
\caption{The first 16 spectral line channel maps of 
H\,I emission with 20.75~km~s$^{-1}$ wide channels
for the central area around NGC~2146. The numbers appearing on the
upper right corner of each channel map are the velocities of the H\,I
in km~s$^{-1}$. The grey areas correspond to H\,I absorption. 
Contours are drawn at -10, -3.5, -2, 2, 3.5, 8, 15, 25 and 40~mJy. 
The peak flux is 46.3~mJy}
\label{fig1}
\end{figure}

\begin{figure}
\vspace{0cm}
\hspace{0cm}\psfig{figure=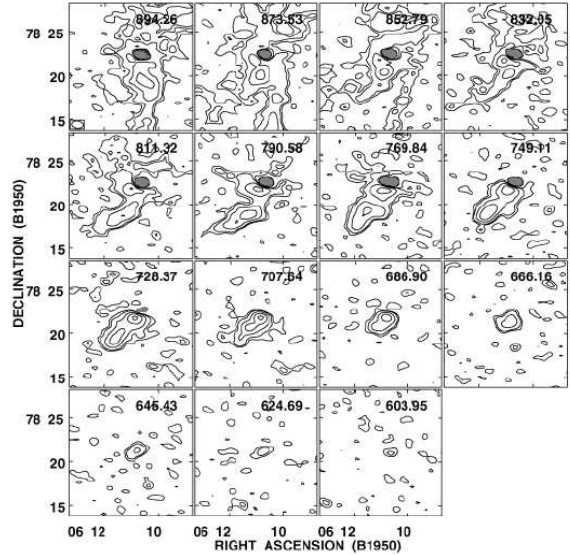,width=8.5cm}
\vspace{0cm}
\caption{The last 15 spectral line channel maps of
H\,I emission with 20.75~km~s$^{-1}$ wide channels
for the central area around NGC~2146. The numbers appearing on the
upper right corner of each channel map are the velocities of the H\,I
in km~s$^{-1}$. The grey areas correspond to H\,I absorption.
Contours are drawn at -10, -3.5, -2, 2, 3.5, 8, 15, 25 and 40~mJy.
The peak flux is 46.3~mJy}
\label{fig2}
\end{figure}

\section{The H\,I distribution}

\begin{figure}
\vspace{0cm}
\hspace{0cm}\psfig{figure=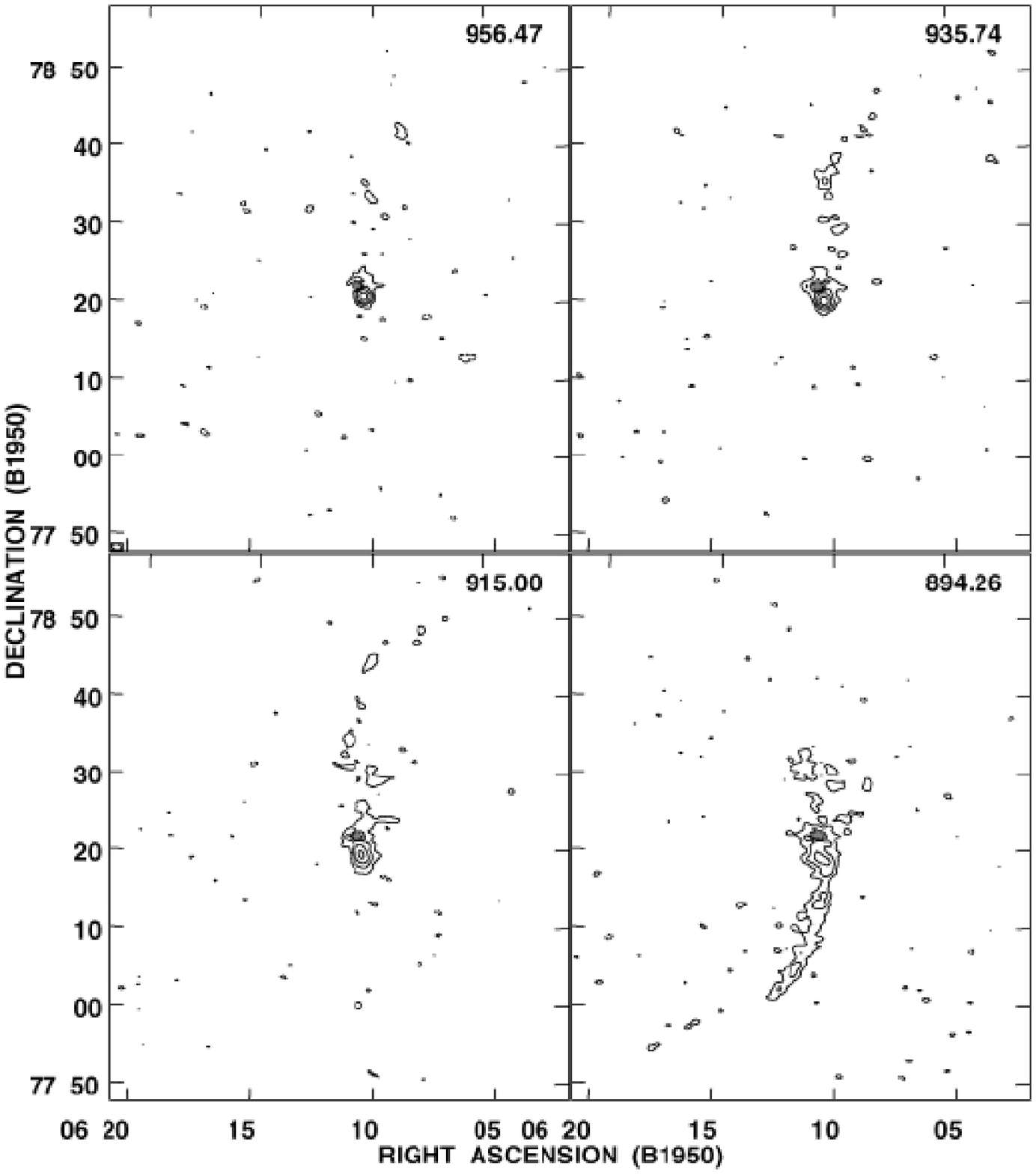,width=8.5cm}
\vspace{0cm}
\caption{H\,I spectral line channel maps showing the extended streams at the 
north and south 
parts of NGC~2146. The grey areas correspond to H\,I absorption.
Contours are drawn at -10, -3.5, 3.5, 8, 15, 25 and 40~mJy.}
\label{fig3}
\end{figure}

\begin{figure}
\vspace{0cm}
\hspace{0cm}\psfig{figure=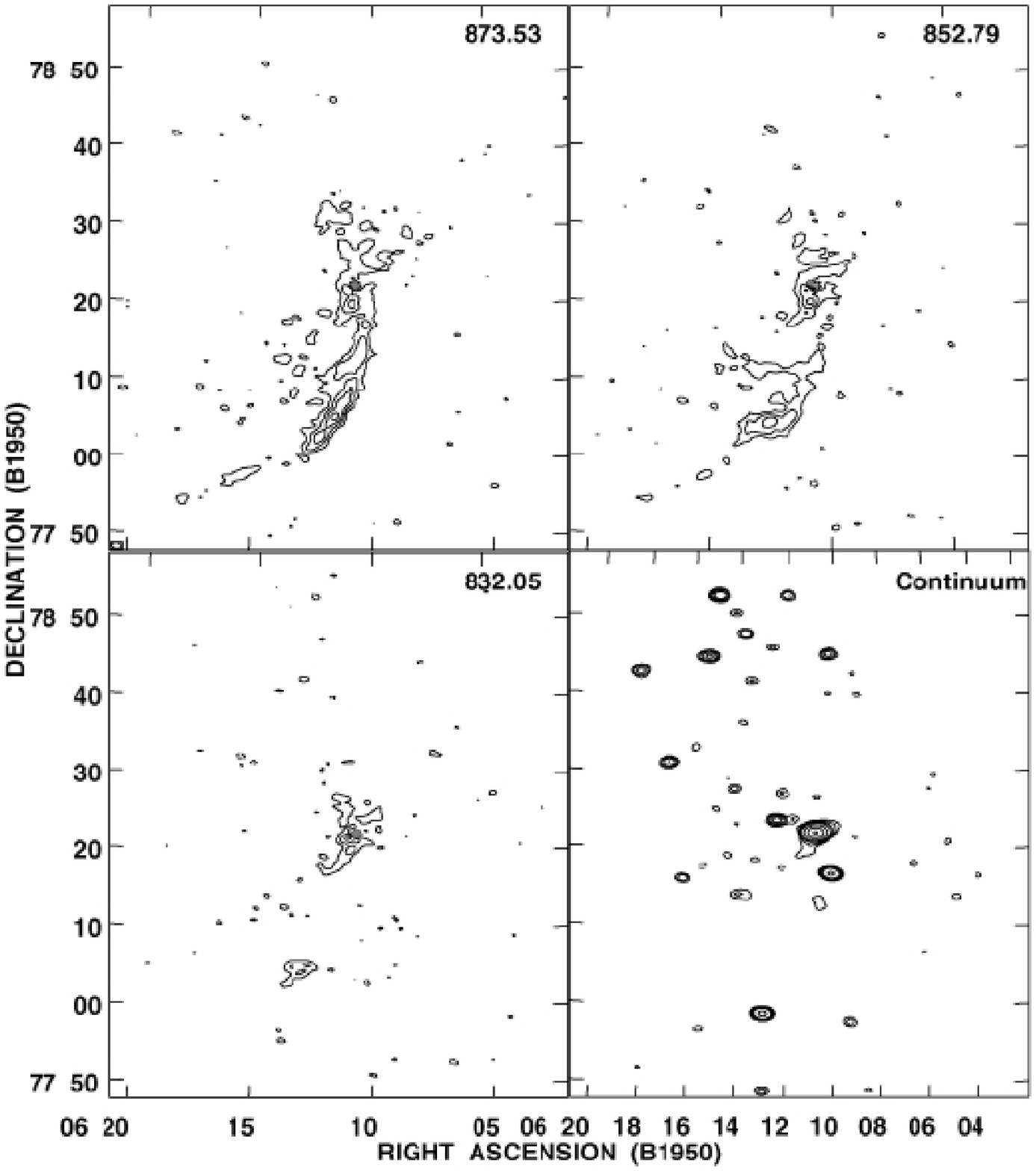,width=8.5cm}
\vspace{0cm}
\caption{H\,I Spectral line channel maps showing the extended streams
at the north and south
parts of NGC~2146 and the continuum map of the area.
The grey areas correspond to H\,I absorption. Contours for the 
channel maps 
are drawn at -10, -3.5, 3.5, 8, 15, 25 and 40~mJy. For the continuum 
the contours are drawn at 2, 4, 8, 16, 64, 256 and 700~mJy, with 
the peak flux being 755~mJy.}
\label{fig4} 
\end{figure}

The channel maps of the H\,I flux of the area around the optical 
galaxy are shown in Figs.~\ref{fig1} and \ref{fig2}. The channels containing emission at large
angular distance from the galaxy, 
together with a map of the radio continuum at 1420MHz, appear in 
Figs.~\ref{fig3} and \ref{fig4}.
These maps reveal elongated streams of neutral hydrogen  towards the
north and the south of the main galaxy, extending out up to 6 Holmberg
radii. This extensive distribution of the H\,I has a ``tail'' 
morphology suggestive of a tidal interaction of NGC~2146 and a companion,
which has not been identified in optical images. 

The southern stream is detected at higher signal to noise ratio than
the northern stream. This occurs because the southern gas appears in
only a few velocity channels, where the spatial structure is dominated
by a single long arc that is nearly coincident in the three main
contributing channels.
There are extensions to this arcing stream,
which, instead of following the general outflow of the H\,I, lie in
smaller arcs of gas that appear to turn eastward and 
separate from the stream, falling back toward the central potential 
of the system.
Further out from the tip of the southern stream appear 
some small H\,I clouds, which do not seem to be gravitationally bound to 
the whole system and may be escaping towards the south east.  

The stream at the north is less prominent, due in part to being
more heavily resolved, both spatially and in velocity, than the
southern stream.
Unlike the nearly constant velocity measured in the south,
there is a prominent velocity gradient across the nothern extended emission
from $\sim$975~km~s$^{-1}$ at position angle $-$15$^\circ$ to 
$\sim$850~km~s$^{-1}$ at position angle $\sim$30$^\circ$. 
Thus, the northern stream is symmetrically placed with respect to 
the southern one, but is not its mirror image.
Instead of being concentrated in only three of our channels 
($\sim$60~km~s$^{-1}$), the northern stream appears to be spread in a fan 
shaped outflow with line--of--sight velocity components ranging from 
$\sim$60~km/s redshifted to $\sim$65~km/s 
blueshifted as a function of position angle on the sky.

\begin{figure}
\vspace{0cm}
\hspace{0cm}\psfig{figure=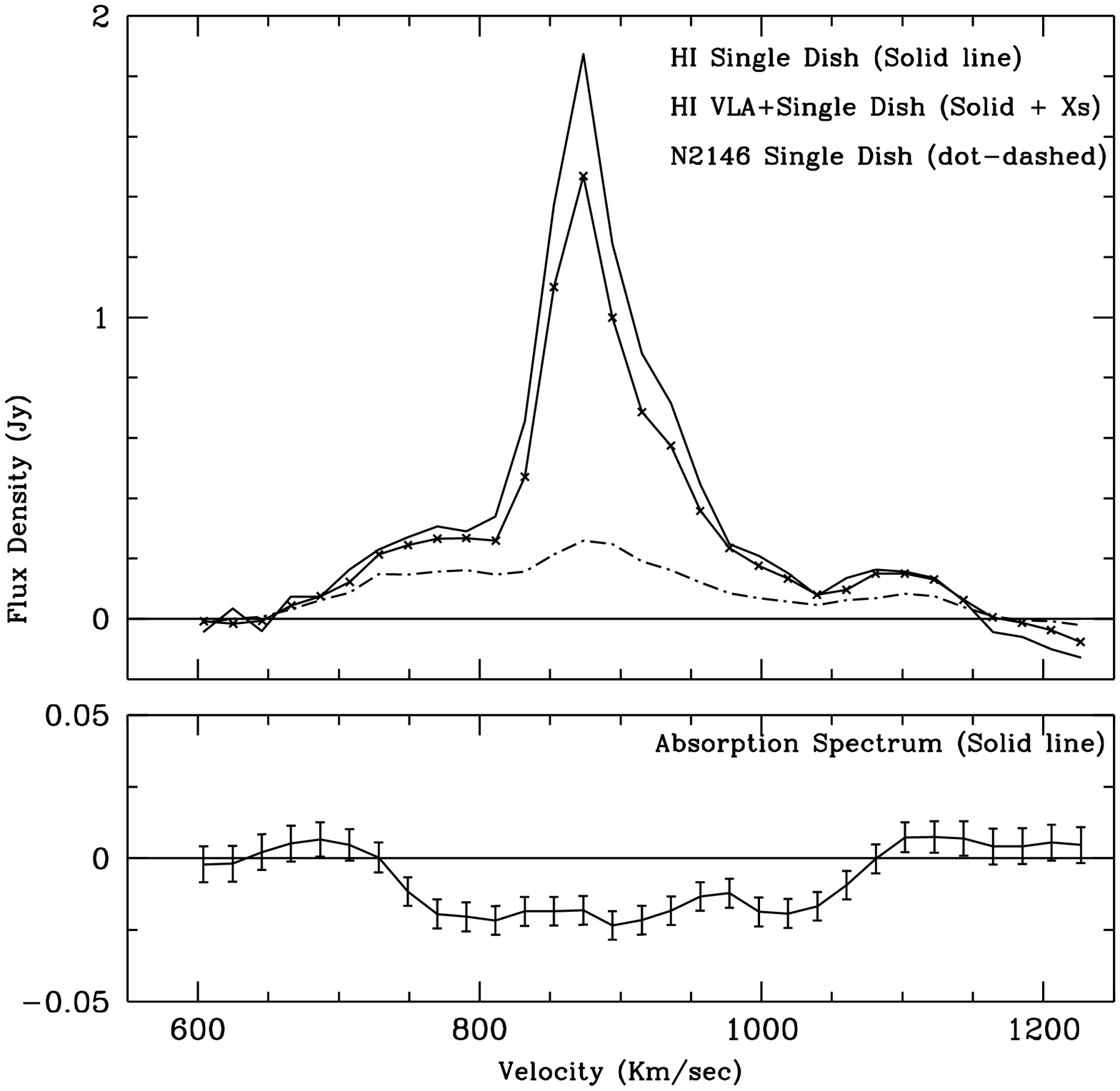,width=8.5cm}
\vspace{0cm}
\caption{The H\,I single dish spectrum of the whole area around NGC~2146, 
and the one obtained after deconvolving the single dish maps with the 
VLA ones to improve the sensitivity. Also is shown the NGC~2146
H\,I profile obtained by Fisher \& Tully (\cite{ft75}) by pointing the 
telescope to 
the optical image of the galaxy (dot dashed line). At the bottom is 
shown the spectrum of the absorption detected by the VLA against the 
radio continuum of NGC~2146 with 5-$\sigma$ errorbars plotted for each spectral 
channel. }
\label{fig5}
\end{figure}

The observed H\,I distribution is consistent with 
that found by Fisher \& Tully (\cite{ft76}), given the 10$^\prime$ beam (FWHM)
of the 91m NRAO telescope. One significant difference is that 
the VLA detects absorption against
the position of the nuclear continuum emission of the galaxy. This absorption 
feature persisted throughout all our reduction, and was consistently 
present in both the VLA maps of the North and the South part of NGC~2146, 
although the position of the galaxy was at a different point in the 
primary beam for each case.
Also, the fact that no deep bowls were seen around the H\,I distribution 
suggests that the absorption is real and not an artifact of 
the interferometer data reduction. 
The spectrum of the absorption is shown in Fig.~\ref{fig5} with 5-$\sigma$ errorbars plotted for each spectral channel. 
This absorption feature would certainly be missed by a single dish telescope 
because of its large beam, in which the absorption is diluted and overwhelmed
by the integrated emission.
Fig.~\ref{fig5} also shows the spectrum of NGC~2146 obtained by 
Fisher \& Tully (\cite{ft76}) by pointing the 91m telescope directly at 
the center of NGC~2146 (dot-dashed line), 
the spectrum of the whole area containing the H\,I cloud made by adding up 
all the flux seen by the NRAO 91-m telescope (solid line), and the spectrum of the total 
H\,I obtained after combining the VLA data with those of the NRAO 91-m 
telescope (solid line with Xs).  Apparently,
we are still missing about 20\% of the 
total H\,I flux; maps produced from the VLA data alone were missing about 
50\% before being
combined   with the single dish data. This deconvolution 
procedure was successful in bringing
up some fine details in the faint extended emission,
especially in the northern stream of gas, where the signal to noise
ratio is low due to being heavily resolved, and also in the
low level emission to the east of the southern arm.

\begin{figure}
\vspace{4cm}
\hspace{6cm}\psfig{figure=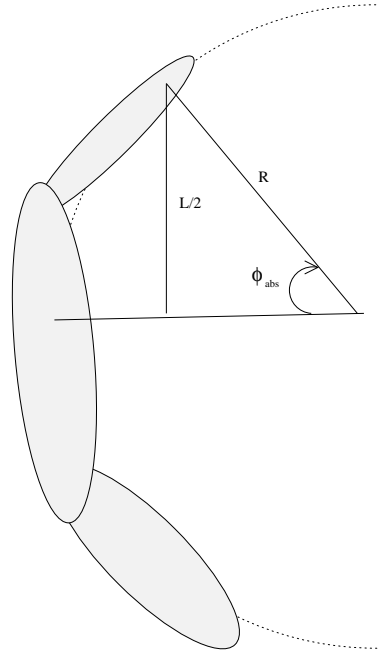,width=5.0cm}
\vspace{-2cm}
\caption{Schematic diagram of the H\,I absorption layer in NGC~2146.}
\label{fig6}
\end{figure}

The absorption is seen against the radio center of the galaxy, which
lies in the region of the optically obscuring dust, and is unresolved
by our observations. It has a velocity width of $\sim$ 350~km~s$^{-1}$
and an average optical depth of 0.03. Assuming a 
spin temperature of 50~K, the measured optical depth
requires an H\,I column density of
$\sim$10$^{21}$~atoms~cm$^{-2}$. The absorption is centered at the
galaxy's systemic velocity and seems to be due to H\,I clouds rotating
together with the rest of the H\,I seen in emission, which happen to lie
in front of the strong radio continuum source of NGC~2146, 4C~78.06. 
However, the large velocity gradient seen accross its 1$^\prime$ span, 
compared to the total width of the
H\,I profile of NGC~2146 ($\sim$~500~km~s$^{-1}$) implies that the
H\,I absorbing layer is not very far from the nucleus of NGC~2146. 
If $V_0$ is the total H\,I profile velocity width of 
NGC~2146 ($\sim$~500~km~s$^{-1}$), and $\Delta v$ is the velocity width of the
absorbing layer seen against the NGC~2146 continuum emission 
($\sim$~350~km~s$^{-1}$) then, assuming a circularly rotating system 
$${{\Delta v} \over {2} } = {{ V_0 } \over {2} } sin\phi_{abs}$$
where $\phi_{abs}$ is shown in Fig.~\ref{fig6}. From Fig.~\ref{fig6} we also see that
$$sin\phi_{abs} = {{ L/2 } \over { R } } $$
where $L$ is the length of the absorbing layer projected on the plane of
the sky ($\sim$ 1 arcsec, or 3.5 kpc; this is an upper limit imposed by
the spatial resolution of our observations),
and $R$ is its distance from the nucleus. Combining these two equations we
estimate an upper limit for the distance of the absorbing layer from 
the continuum background to be
$$R = {{ L~V_0 } \over { 2~\Delta v } } = 2.5~~{\rm kpc} $$

\begin{figure}
\vspace{0cm}
\hspace{0cm}\psfig{figure=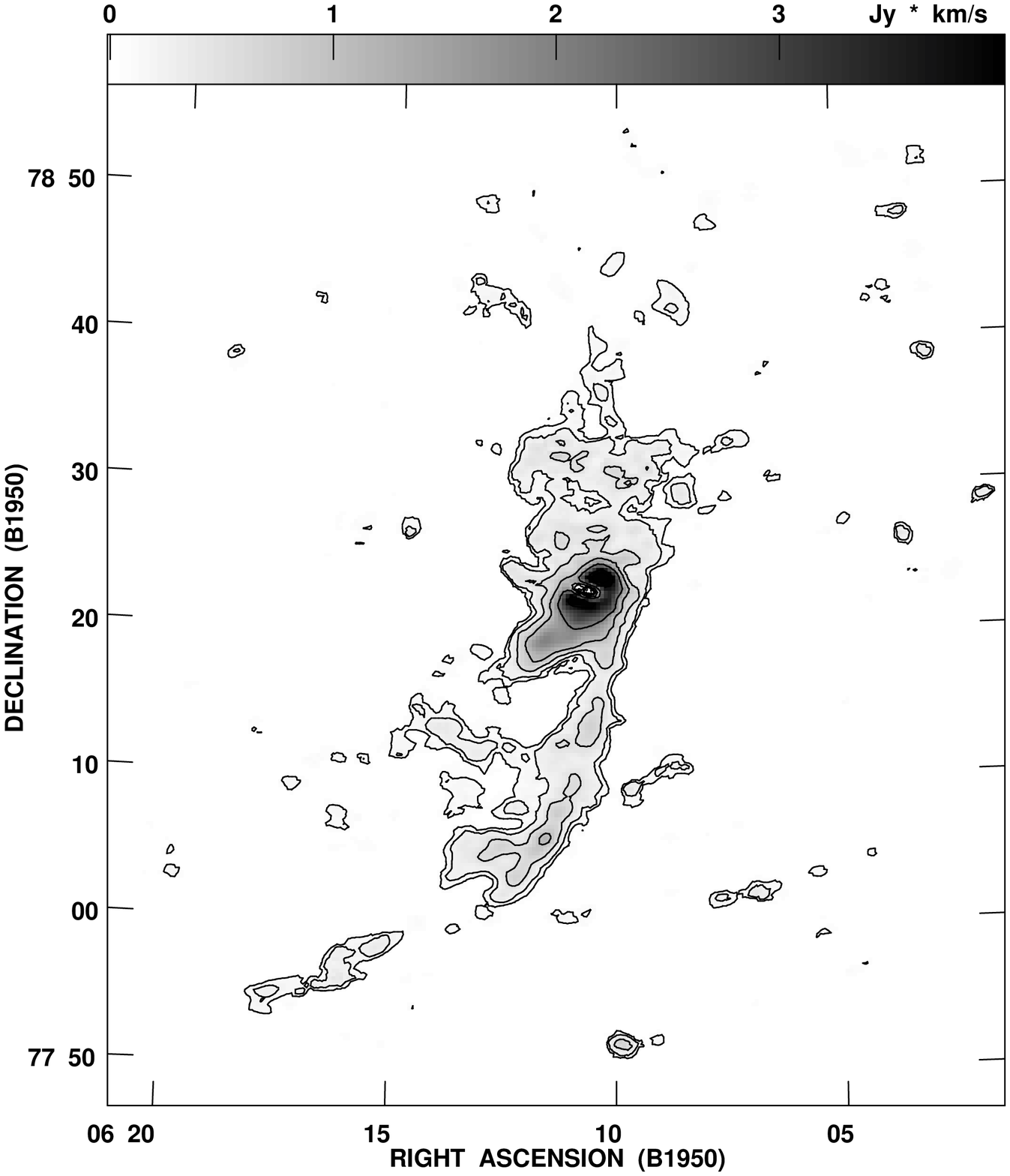,width=8.5cm}
\vspace{0cm}
\caption{The total integrated H\,I flux density of NGC~2146. Contours 
are drawn 
at -0.5, -0.2, -0.1, 0.1, 0.2, 0.5, 1, 2, 4 and 5~Jy~km~s$^{-1}$. 
The peak flux is 5.8~Jy~km~s$^{-1}$.}
\label{fig7}
\end{figure}

\begin{figure}
\vspace{0cm}
\hspace{0cm}\psfig{figure=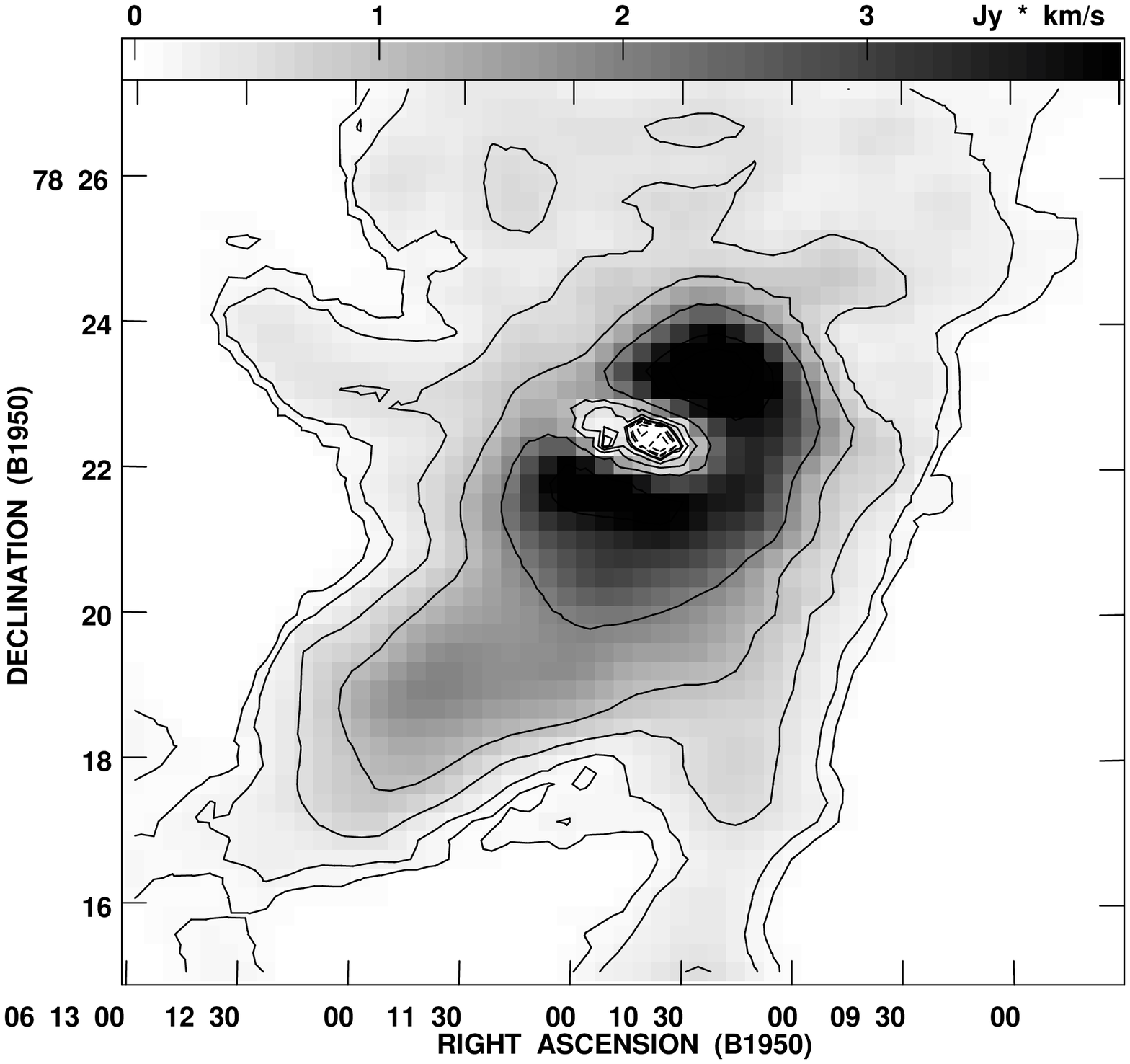,width=8.5cm}
\vspace{0cm}
\caption{The total integrated H\,I flux density of NGC~2146. Contours
are drawn 
at -0.5, -0.2, -0.1, 0.1, 0.2, 0.5, 1, 2, 4 and 5~Jy~km~s$^{-1}$. 
The peak flux is 5.8~Jy~km~s$^{-1}$.}
\label{fig8}
\end{figure}

The integrated H\,I flux density map is shown in
Figs.~\ref{fig7} (large scale) and \ref{fig8} (small scale). We estimate 
the total mass for the cloud to be about $6.2\times
10^9~M_\odot$ ($H_0$=75~km~s$^{-1}$~Mpc$^{-1}$)
of which $1.6\times10^9~M_\odot$ 
come from the bright region around the galaxy itself, and 
$4.6\times10^9~M_\odot$  from the extended component. 
Furthermore, the extended distribution is not symmetric with respect
to the amount of gas in the south ($3.1\times10^9~M_\odot$)
and the north ($1.5\times10^9~M_\odot$). 
For the above calculations a velocity component of 
223~km~s$^{-1}$ of our Sun towards NGC~2146's direction has been assumed, in 
order to refer the redshift velocity to the center of mass of the Local 
Group. 

\begin{figure}
\vspace{0cm}
\hspace{0cm}\psfig{figure=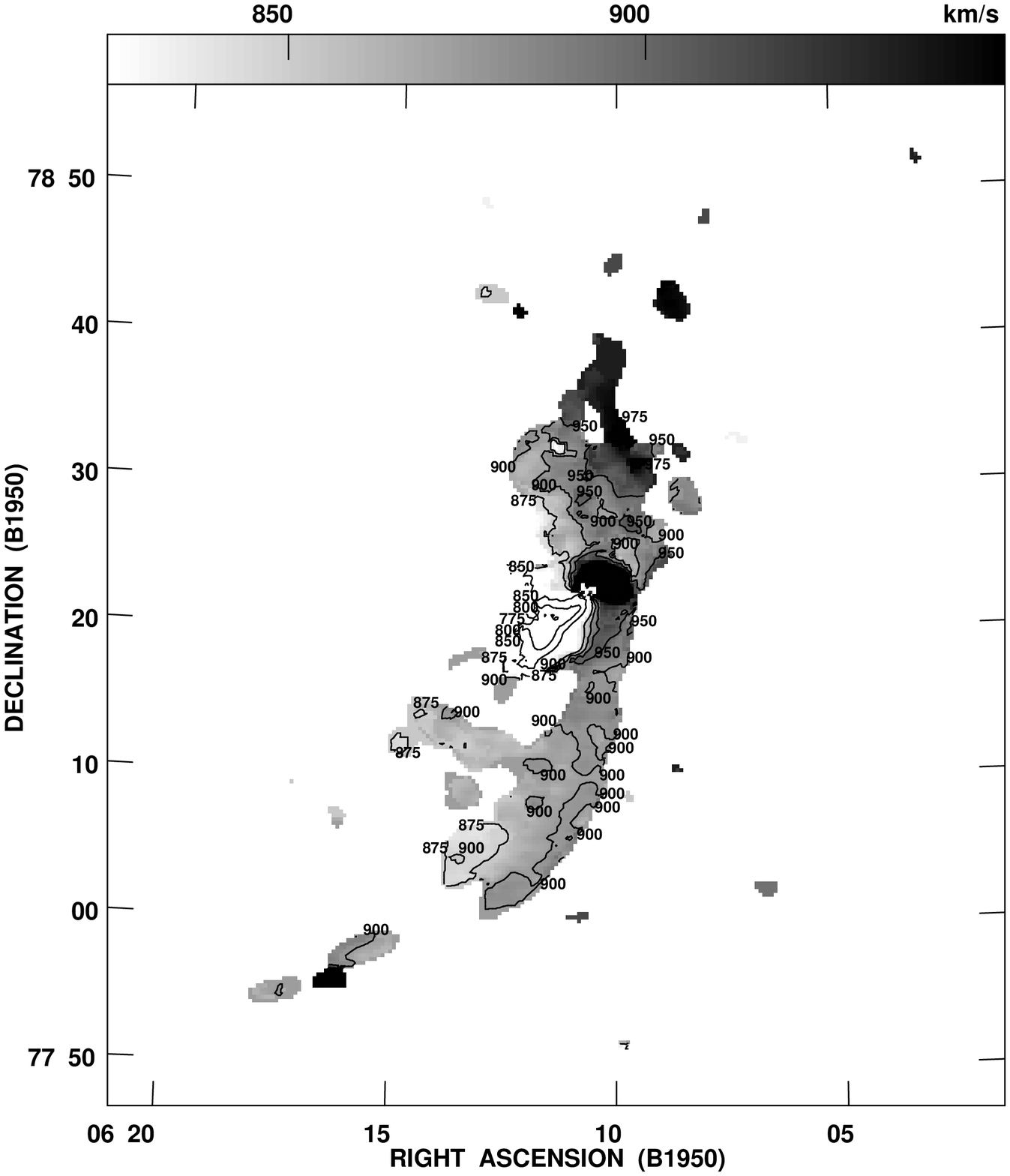,width=8.5cm}
\vspace{0cm}
\caption{The H\,I intensity weighted mean radial velocity field of NGC~2146.
Contours are drawn at 750, 775, 800, 850, 875, 900, 950, 
975, 1000, 1050 and 1100~Km~s$^{-1}$. The greatest velocity is 
1201~km~s$^{-1}$. }
\label{fig9}
\end{figure}
 
\begin{figure}
\vspace{0cm}
\hspace{0cm}\psfig{figure=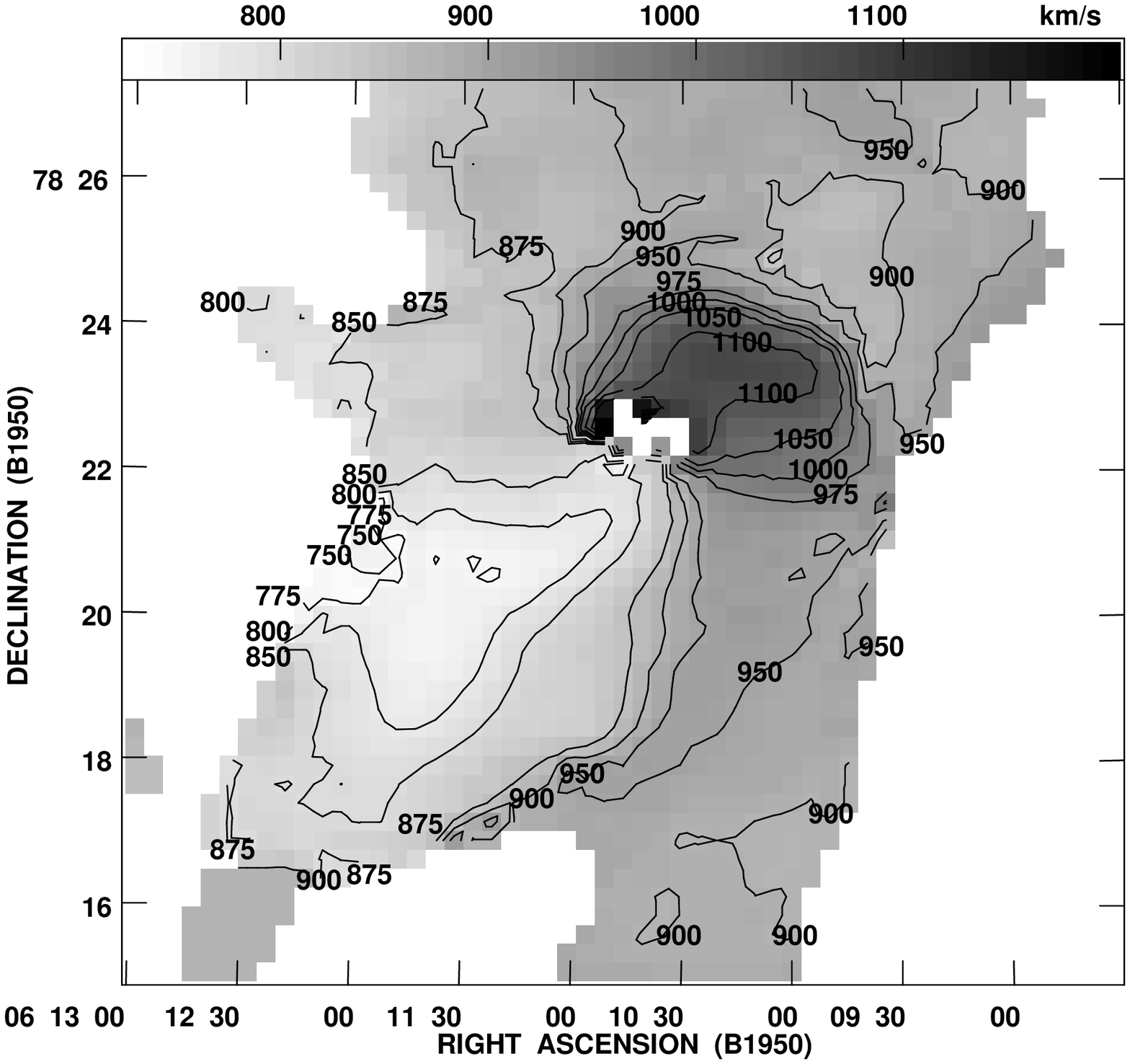,width=8.5cm}
\vspace{0cm}
\caption{The H\,I intensity weighted mean radial velocity field in the 
central region of NGC 2146. 
Contours are drawn at 750, 775, 800, 850, 875, 900, 950,          
975, 1000, 1050 and 1100 Km s$^{-1}$. The greatest velocity is 1201
km s$^{-1}$.}
\label{fig10} 
\end{figure}

Figs.~\ref{fig9} and \ref{fig10} show the intensity weighted mean radial velocity field
on both large and small scale. Points with negative flux due to absorption
have been excluded from this calculation, and
only emission brighter than 1.3~mJy has been taken into account. 
It is clear from the velocity field that the main galaxy has a   
differentially 
rotating disk with characteristic rotational speed of about 
250~km~s$^{-1}$. The extended H\,I disk, however, appears severly lopsided 
as can immediately be
seen from Fig.~\ref{fig10} since the galaxy is seen almost edge on.

The H\,I disk of the main galaxy has a well defined plane of rotation, 
with an inclination nearly perpendicular to the sky plane. The extended 
H\,I stream to the south, however, has very little velocity gradient
along its entire length, with an average velocity that is coincident
with the systemic velocity of the central galaxy.  This situation
occurs when the outlying stream and the galaxy centroid lie in the
plane of the sky and implies that we observe the southern
stream from a preferred viewing angle, in which it appears laid out in the 
probable plane of the interaction, which must have been very nearly
perpendicular to the disk plane of NGC~2146. The northern stream, on the
other hand, does show a significant velocity gradient across it, 
and the orientation of the stream cannot be inferred in a 
model--independent way.

\begin{figure}
\vspace{0cm}
\hspace{0cm}\psfig{figure=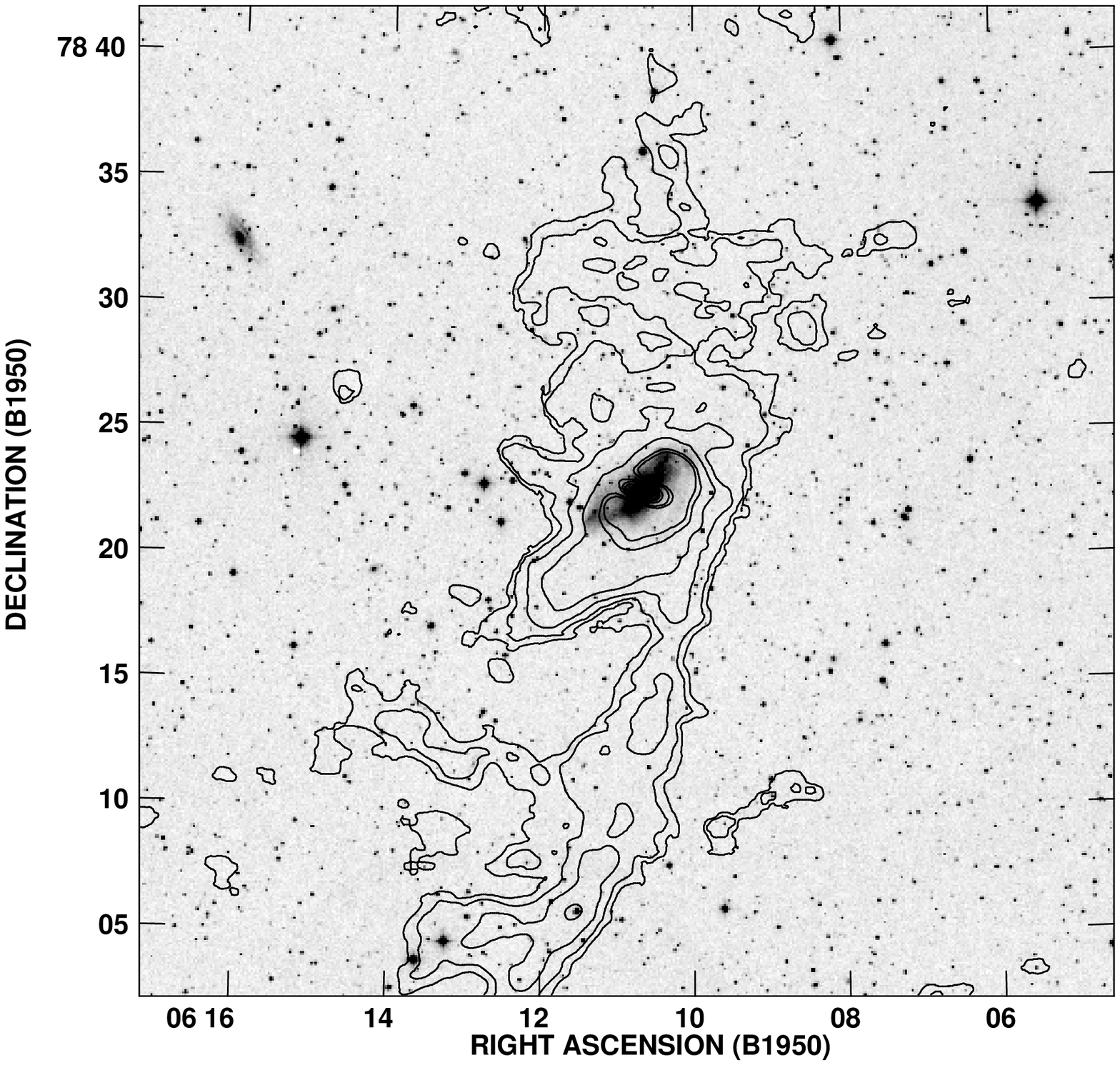,width=8.5cm}
\vspace{0cm}
\caption{The total integrated H\,I flux density of NGC~2146 superimposed on
an optical image of the galaxy from the Digitized Sky Survey (DSS1).
The contours drawn are the same as in Fig.~\ref{fig7}.}
\label{optfig}
\end{figure}

In Fig.~\ref{optfig} the total integrated H\,I flux density of NGC~2146 is
plotted superimposed on an optical image of the galaxy from the Digitized Sky
Survey (DSS1). The morphology of the H\,I cloud does not show any preferential 
alignment with respect to the direction of NGC~2146A, and therefore shows no 
evidence for interaction between NGC~2146 and NGC~2146A.
The faint arm extending southwards from the top NW
corner of the optical image is in the general direction of the H\,I southern 
stream. However, the velocity gradient of H\,II regions of 
$\sim$100 km sec$^{-1}$ observed
by Young et~al. (\cite{young}) across this feature is 
incompatible with the absence of a velocity gradient in the southern H\,I 
stream and its alignment with the plane of the sky. It seems thus that 
this arm is part of the main galaxy which was disturbed from an interaction but
still remained under the strong gravitational influence of NGC~2146. The 
loop structure at the SE of the optical image of the galaxy shows  
higher velocities than those expected if it was rotating in the plane of the 
galaxy (Young et~al., \cite{young}) presenting further evidence for a 
tidal interaction. However, there
is no apparent correspondence with any H\,I feature. 

\section{Discussion}

The complex appearance of this system may represent the aftermath of
an encounter between NGC~2146 and a slightly less massive but gas--rich
galaxy, probably a late--type LSB spiral, with a slowly rising rotation
curve indicating little or no bulge (de Blok et~al. \cite{deblok}). 
Numerical simulations by 
Wallin \& Stuart (\cite{wallin}) indicate that the outcome of an interaction
depends crucially on the orientation of each galaxy's rotational angular 
momentum vector relative to the plane of the interaction. If the 
rotational angular
momentum vector lies in the plane of the interaction, the galaxy
comes through with very little loss of mass. On the other hand, if the
rotational angular momentum aligns with the orbital angular momentum,
a large fraction of the galaxy is stripped away. 

\begin{figure}
\vspace{-2cm}
\hspace{0cm}\psfig{figure=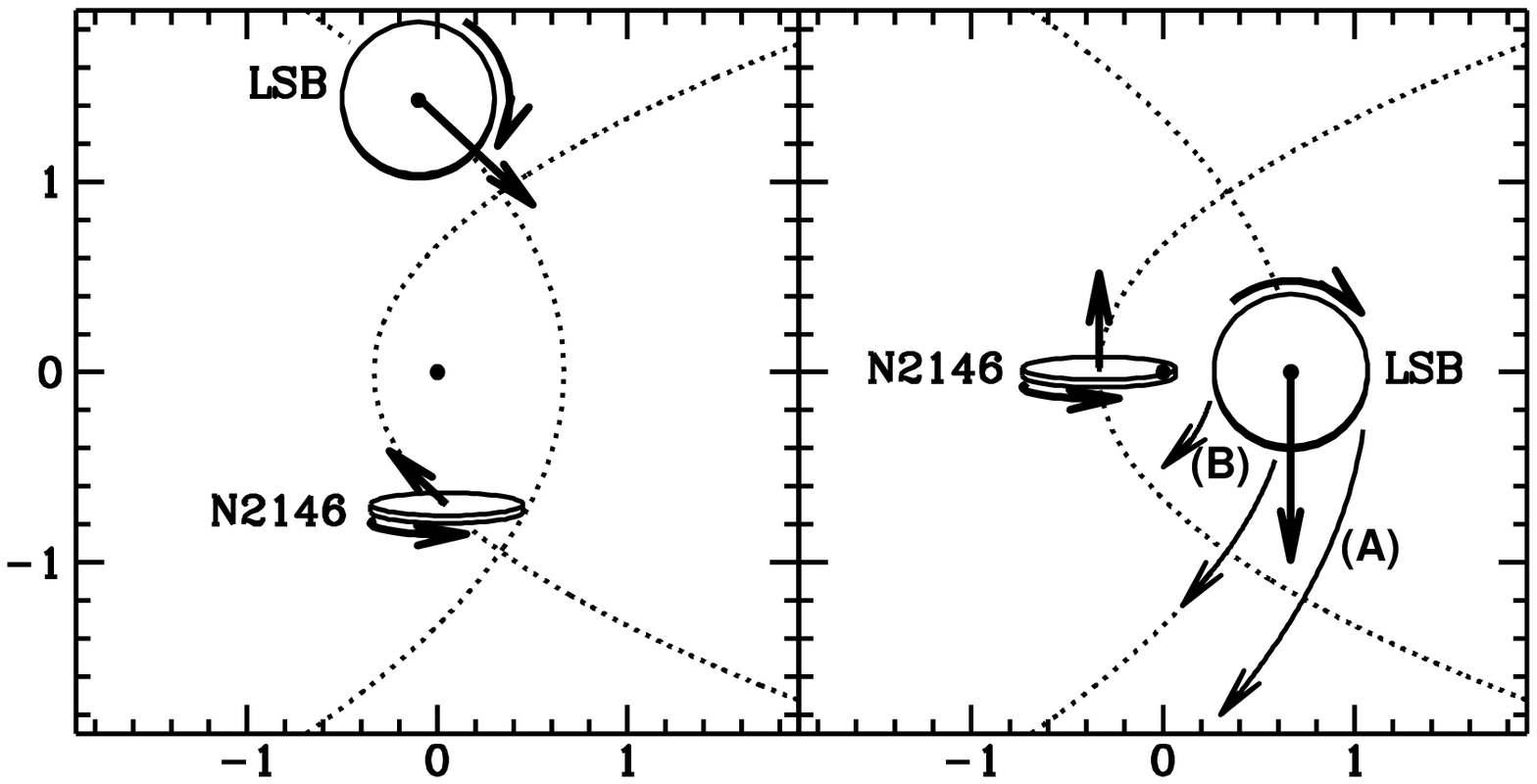,width=8.5cm}
\vspace{-2cm}
\caption{ Schematic of Interaction. {\it Left Panel} Pre-encounter.
{\it Right Panel} Closest approach. The schematic shows 
the observer's view of an interacting system where the translation 
of the incident galaxies takes place in the sky plane. The rotational
angular momentum vector for NGC 2146 galaxy lies in the sky plane, and
the rotational angular momentum vector for an incident LSB galaxy
is oriented perpendical to the sky plane. }
\label{fig11}
\end{figure}

NGC~2146 may have interacted with a gas--rich ``intruder'' whose rotational
angular momentum vector was oriented perpendicular to that of NGC~2146 and 
perpendicular to the plane of the interaction, which in this case is close to
being coincident with the plane of the sky. Fig.~\ref{fig11} is a schematic
diagram of the encounter.
In this configuration, NGC~2146 has a rotational
angular momentum vector lying in the interaction plane and is able to
preserve its identity as a rotating disk galaxy, while the same
encounter essentially destroys the intruder.

The numerical simulations of Wallin \& Stuart (\cite{wallin}) predict in such cases
that the mass fraction transfered from the intruder to the main galaxy
can be as large as 0.5, and the mass fraction lost by the intruder
that eventually escapes
from the whole system can reach up to 0.2. Wallin \& Stuart
used a model with restricted 3--body interactions (two large point
masses binding massless test particles), and more realistic models
with distributed mass are likely to suffer even greater destruction
for the case of aligned orbital and rotational anglular momentum, as is
indicated by some test cases run for us by J. Gerritsen (private communication)
using his tree-code implementation (Gerritsen \cite{gerritsen}). 

The outcome of the tidal interactions is the development of gas streams
on opposite sides of both galaxies. The velocity gradient of the gas in 
the northern stream is consistent with the sense of rotation of NGC~2146, 
which may indicate that this gas originated from the H\,I disk of NGC~2146. 
The gas at the southern stream, however, appears only around the 
systemic velocity of the system, and is likely to be the
tidally dispersed remnant of the intruder.
As the galaxies swirled around each other, H\,I clouds released from the
intruder's side opposite to NGC~2146 were 
boosted to escape velocity by the combination of
the intruder's translational and rotational velocities. In this
scenario, the outermost clouds of the intruder experience the gravitational
slingshot effect familiar in interplanetary
spacecraft trajectories to reach the
outer solar system.
We next test the plausibility of such an encounter by exploring the
timing arguments for the formation and lifetime of the
southern plume near NGC~2146.  For our purposes, we approximate
the relative speeds of the masses using the vis viva equation
to describe the relative speed $V$ of the centers of mass for
the two masses, $M_{NGC2146}$ and
$M_{intr}$, as a function of their separation $R$ and $a$, the
semimajor axis of a possible elliptical orbit of the intruder about NGC~2146.
Then,
\begin{eqnarray}
V^2 &=& G(M_{NGC2146}+M_{intr})\left[\frac{2}{R}-\frac{1}{a}\right]
\nonumber\\
&=& GM_{NGC2146}\left(1+\frac{M_{intr}}{M_{N2146}}\right)\left[\frac{2}{R}\right] 
\nonumber\\
&\approx& (350 {\rm ~km~s}^{-1})^2
  \left(1+\frac{M_{intr}}{M_{NGC2146}}\right)\left[\frac{15 {\rm kpc}}{R}\right]
  \nonumber
  \end{eqnarray}
where we have made the assumption that the intruder falls in from
large distance so that $a\rightarrow\infty$ and that the mass of
NGC~2146 can be approximated by adopting the rotational
velocities of $V_o\simeq$250~km~s$^{-1}$ that are measured
in the outskirts of NGC~2146 at $R_o=$15~kpc to obtain
$GM_{NGC2146}\approx R_oV_o^2$.

For purposes of this illustration, we consider an HI-rich,
``LSB'' intruder, with $M_{intr}=M_{NGC2146}/4$ and $R_{intr}\approx R_o$,
having a gradually rising rotation curve that reaches
$V_{rot}\approx 125$~km~s$^{-1}$ at the outer edge of the galaxy's
disk.  Under these assumptions, the relative speeds of the centers
of mass for the two galaxies will be $V\approx 277$~km~s$^{-1}$ for
a grazing encounter where $R=30$~kpc.  Four-fifths of the encounter
speed (${\sim}$222~km~s$^{-1}$) is carried by the  center of mass of
this intruder at a distance of 24~kpc from the system barycenter.
By the time of closest approach, the intruder will have already
become tidally distorted, and as it passes pericenter, simple tidal force
arguments show that it will be unbound throughout its disk, provided
it has a slowly rising rotation curve. However,
in order to estimate the ``launch speed'' of debris in the
outskirts of the intruder (trajectory A in Figure 12), 
we consider the effect of adding 
the $V_{rot}$ of the intruder to the speed of its
center of mass.

Under these assumptions, gas clouds at the far side of the intruder
from NGC~2146 are traveling at $\sim$346~km~s$^{-1}$ at a distance
of 39~kpc from the barycenter.  We estimate the escape velocity from
this distance by lumping $M_{NGC2146}+M_{intr}$ at the barycenter
to find $V_{esc}\approx 240$~km~s$^{-1}$, which is roughly 100~km~s$^{-1}$
less than the speed of the outer gas clouds. Once the process of
dismantling the intruder begins, the mass lost lowers the
binding energy, allowing the intruder to be shredded throughout.

Gas clouds on the inside of the intruder closest to the system
barycenter, of course, find that whatever rotational component
of velocity they still carry is counter to the orbit around
NGC~2146.  The dynamics in this region are clearly complicated,
so that subtracting the 125~km~s$^{-1}$ rotational speed from the
222~km~s$^{-1}$ intruder center of mass speed to leave 
$\sim$100~km~s$^{-1}$ forms only a rough guess.  However, the implication
is still that the inner gas is progressing in the same direction as the 
bulk of the material from the dismemberment of the intruder.

As a result, we picture the southern HI plume to be the result of
tidal shredding of a high angular momentum LSB galaxy.  The test
simulation by Gerritsen (private communication) confirms that this
could be accomplished without enhancing star formation in the
intruder.

In order to make some statement about the time span since closest
approach, we make the following simplifying assumptions:  First, the
high speed clouds from the outskirts of the intruder escape
(trajectory A in Figure 12) from
the system by overcoming the gravitation potential created by the
the sum of the two original masses, concentrated at the system
barycenter.  Second, the remnants such as the HI spur running
nearly east-west at declination ${\sim}78^{\circ}12'$ find their
acceleration dominated by the mass of NGC~2146 alone. If we assume
that the gas in this spur is located at the turn-around point of
its trajectory, we can estimate the major axis of its orbit and
deduce the time span required for it to return to closest approach
to NGC~2146. The spur falls ${\sim}14'$ or 49~kpc from the center of
NGC~2146. Assuming these clouds were launched from a distance of
${\sim}25$~kpc on the opposite side of NGC~2146, this trajectory (B in 
Figure 12)
can be approximated by an ellipse of major axis $2a=74$~kpc.
Using the rotational
velocity  $V_o=\sim$250~km~s$^{-1}$ for NGC~2146 at $R_o=$15~kpc,
we deduce that the orbital period $T(a)$ would be 
$(T/0.4{\rm Gyr})^2= (a/15{\rm kpc})^3$ as a function of $a$.
Thus, half of the period for $a=37$~kpc is $\sim$0.8~Gyr.
This forms an estimate of the time elapsed since the closest 
approach.

The gas clouds ejected along trajectory A have reached
${\sim}50'$ or 180~kpc during 0.8~Gyr, for an average speed of
225~km~s$^{-1}$. If the clouds were launched with the
346~km~s$^{-1}$ estimated above, then they will have decelerated
along their path, consistent with this estimate for average speed.

In this scenario, the central regions of the destroyed intruder are likely
to lie somewhere along the south stream. 
Indeed, there is a concentration of gas, located at declination
78$^\circ$04$^\prime$, where the south stream has a kink (most clearly
seen in channels with velocities of 853 and 832~km~s$^{-1}$ in
Fig.~\ref{fig4}). The concentration 
has galaxy sized dimensions and a velocity gradient of 60~km~s$^{-1}$
across its 30 kpc extent, 
contains $\sim$~1.5$\times$10$^8~M_\odot$ in neutral hydrogen, and has an
H\,I column density of $\sim$~2$\times$10$^{20}$~atoms~cm$^{-2}$, 
which is high enough to trigger star formation that might be
detectable in deep CCD images. All these make this object a very good
candidate of being a remnant of the long sought companion of NGC~2146.
Clearly more observations need to be undertaken to clarify the nature
of this object, and test the validity of this scenario.

Another plausible scenario for this system is that it is at the final stage of
merging and the small companion has been completely stripped off of its gas
leaving no sign of its nucleus, as suggested by Hutchings et~al. (\cite{hutchings}) and 
Lisenfeld et~al. (\cite{lisenfeld}). In view of the support that numerical simulations can 
provide to such mergers
(Barnes \cite{barnes}) and if the above mentioned H\,I concentration at the south 
does not provide evidence for a companion nucleus around NGC~2146 this seems
also a very attractive possibility.

The NGC~2146 system provides another example to add to those of Hibbard \& 
van Gorkom (\cite{hibbard}) where 
galaxy-galaxy interactions inject galactic gas into galactic
halo regions, as well as ejecting gas with sufficient velocity that it
can escape
to the intergalactic medium. Depending on the mass ratio of the interacting
galaxies, the relative inclinations of their disks and the impact parameter
of the encounter the mass lost in the interaction 
can be up to 60\% of the mass of the companion galaxy (Wallin \&
Stuart \cite{wallin}). This gas contributes to the enrichment of the intergalactic
medium in metals and if such events had been very common at some earlier
epochs they may help towards finding the objects responsible for the
metal absorption-line systems seen in abundance in QSO spectra.

Which of the above scenaria is correct remains yet undetermined. Numerical
simulations based on the morphological and kinematical information
presented may give us a clearer picture of how the system looked a billion
years ago. Nevertheless it is striking that the final result of
this collision is a morphologically classified
spiral galaxy, although both the galaxy NGC~2146 and its surrounding
H\,I cloud seem
to have contained approximately equal amounts of neutral hydrogen.
This system does provide evidence that mergers of two significant systems
can be important events in the history of spiral as well as elliptical
galaxies.

The encounter will affect the host galaxy for a long time into the
future, since the outlying gas residue is likely to fall
into the galaxy both from north and south meeting the rotational plane of
the galaxy at a a significant angle of inclination to its
internal orbital plane; this may provide a trickle of
missaligned angular momentum for a long time as gas on different
trajectories turns around and falls back. In this sense, the effect of
this collisional debris
will resemble Binney's ``Cosmic Drizzle'' (\cite{binney}) as a mechanism 
for creating long--lived warps in large, isolated galaxies.

\section{Conclusions}

By combining single dish observations and D~-~configuration VLA data we have 
produced sensitive, high quality images of the starburst galaxy NGC~2146. 
These images reveal elongated streams of neutral hydrogen extending out up to 
6 Holmberg radii towards both the north and the south of the main galaxy. 

The H\,I morphological and kinematical picture suggests that this galaxy 
underwent a strong interaction with a LSB companion, which was destroyed 
during the encounter. Although quite disturbed, NGC~2146 retained the 
morphological characteristics of a spiral galaxy. Analysis of the trajectory 
of the outlying gas suggests that the interaction was in its most violent 
phase about 0.8$\times$10$^9$ years ago. The infall of the outlying gas to 
NGC~2146 will continue for a similar timespan preserving a long--lived warp 
in the disk of this galaxy. 

\begin{acknowledgements}
We wish to acknowledge the National Astronomy and Ionosphere Center at
the Arecibo Observatory for their hospitality during part of the data 
reduction. The National Radio Astronomy Observatory (NRAO) is
operated by Associated Universities, Inc. under cooperative agreement with
the National Science Foundation. This research has made use of the NASA/IPAC
Extragalactic Database (NED) which is operated by the Jet Propulsion
Laboratory, Caltech, under contract with the National Aeronautics 
ans Space Administration. This work has been supported by 
National Science Foundation Grant AST 91-19930. 
\end{acknowledgements}

\end{document}